\documentstyle[version2,aps]{revtex}
\begin{document}
\draft
\preprint{DA - 07}
\twocolumn[\hsize\textwidth\columnwidth\hsize\csname
  @twocolumnfalse\endcsname
\begin{title}
\begin{center}
{\bf A simple scaling law between the total energy of a free atom \\
and its atomic number}
\end{center}
\end{title}
\author{W.~T.~Geng}
\begin{instit}
Department of Physics \& Astronomy, Northwestern University, Evanston,
IL 60208
\end{instit}
\begin{abstract}
A simple, approximate relation is found between the total energy of a free atom and
its atomic number: $E\simeq -Z^{2.411}$. The existence of this index is
inherent in the Coulomb and many-body nature of the electron-electron interaction
in the atomic system and cannot be fabricated from the existing fundamental 
physical constants.
\end{abstract}
\vskip2pc]
 In a recent work on the calculation of the cohesive energy of elemental
 crystals,\cite{wtg} we have calculated the total energy for all atoms with 
 Z$\leq$92. Our calculations were based on the density functional theory 
 with the local density approximation.\cite{DFT}
 Parameterization
 of the exchange-correlation interaction is that of Hedin-Lundqvist.\cite{LDA}  For the first time, 
 we plotted the total
 energy ($E$) versus the atomic number ($Z$) curve (Fig. 1), in an attempt to 
 gain some physical insight into the density functional 
 theory. Surprisingly, it is found that this curve can be very well fitted by a 
 scaling law

 $$E=-Z^n, \ \ n=2.411 $$

 To make the $E \sim Z$ relation more illustrative, the $n \sim Z $ curve
 is plotted in Fig. 2 (down triangles)). The power index $n$ is almost 
 constant (close to 2.41) for atoms with 
 $4\leq Z \leq 92$. If there is no interaction between 
 electrons, $n=3$; and if there is only one electron outside this nucleus, $n=2$.
 Since the electron-electron interaction increases the total energy (i.e., less
 negative), $n$ should meet $2<n<3$.
 The existence of such a near-constant power index is astounding because, due to
 the complexity of the quantum many-body problem, it's never
 been expected that the total energy of an atom other than hydrogen should have so
 simple a relation with its atomic number. 

 Exceptions occur in the cases of hydrgen, helium, and lithium. For hydrogen,
 $Z=1$ and $E=-0.976~Ry$, therefore, $n$ has no definite value. For helium and 
 lithium, $n$=2.506 and 2.447, respectively, apparently larger than 2.41.
 Experimental data\cite{EXPT}, which is non-relativistic and available up to
 argon, are denoted by triangles in Fig.2. Also listed are the calculated power
 index from Desclaux's Hartree-Fock atomic total energy data.\cite{HF} Open circles
 represent non-relativistic treatment and solid circles denote relativistic 
 treatment. It is seen that all these four groups of $n$ have values with very 
 limited diviation. It's then concluded that the scaling law is not an outcome
 of the density functional theory, where both the exchange and correlation 
 interactions are considered, nor a result of the Hartree-Fock approximation, in 
 which only the exchange interaction is counted. Although relativistic effects makea 
 difference in the index $n$, the approximate scaling law holds for both cases.

 As ionization potentials show very strong effects of chemical periodicity, it is
 of much interest to see whether they exert a periodic effect on the atomic total 
 energy too. We replot the $n\sim Z$ curves in Fig.3, a higher resolution graph.
 $n$ shows apparent oscillatory behavior for atoms lighter than krypton.
 But for heavier atoms, it displays monotonic character. This is due to the fact
 that the ionization potentials are so small as to be averaged out for the heavy 
 atoms. It is worth noting that the solid circle denoting Desclaux's 
 relativistic francium falls out of the otherwise smooth curve. There must be an 
 abrupt mistake, probably a typo, in the reported total atomic total energy of
 francium. 
   
   From the comparisons between density functional theory and Hartree-Fock approximation,
   relativistic and non-relativistic treatments, we can conclude that the existence 
   of such a simple relation between the total energy of an atom and its atomic number
   is independent of the framework in which the calculations of atomic total energy 
   are carried out. It is inherent in the Coulomb and many-body nature of the atomic 
   system. Apparently, this power index cannot be fabricated from the existent
   fundamental physical constants such as $\hbar$, $c$, $e$, etc., and can only be 
   built into a new many-body quantum theory.


\acknowledgments 
The author acknowledges helpful
discussions with Professors Ding-Sheng Wang and A. J. Freeman.

\figure{Atomic total energy (Ry) given by density functional theory.
}

\figure{Calculated power index $n$ in relation $E=-Z^n$. Down triangles 
 are our results from density functional theory; open (non-relativistic) and 
  solid (relativistic) circles
   are calculated from Desclaux's Hartree-Fock data; triangles denote experimental
    data (non-relativistic).}

\figure{A replot of Fig.2 with higher resolution.}
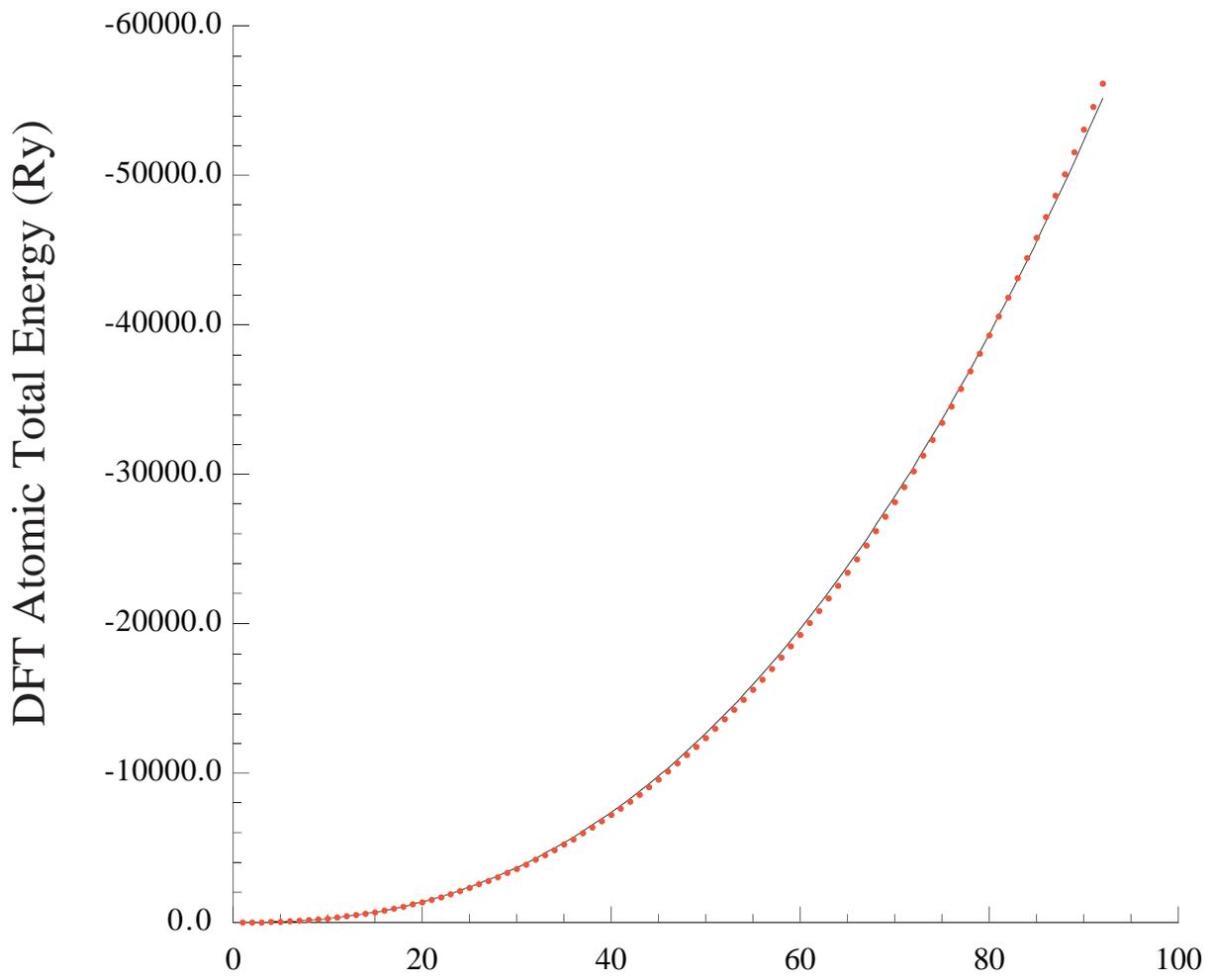
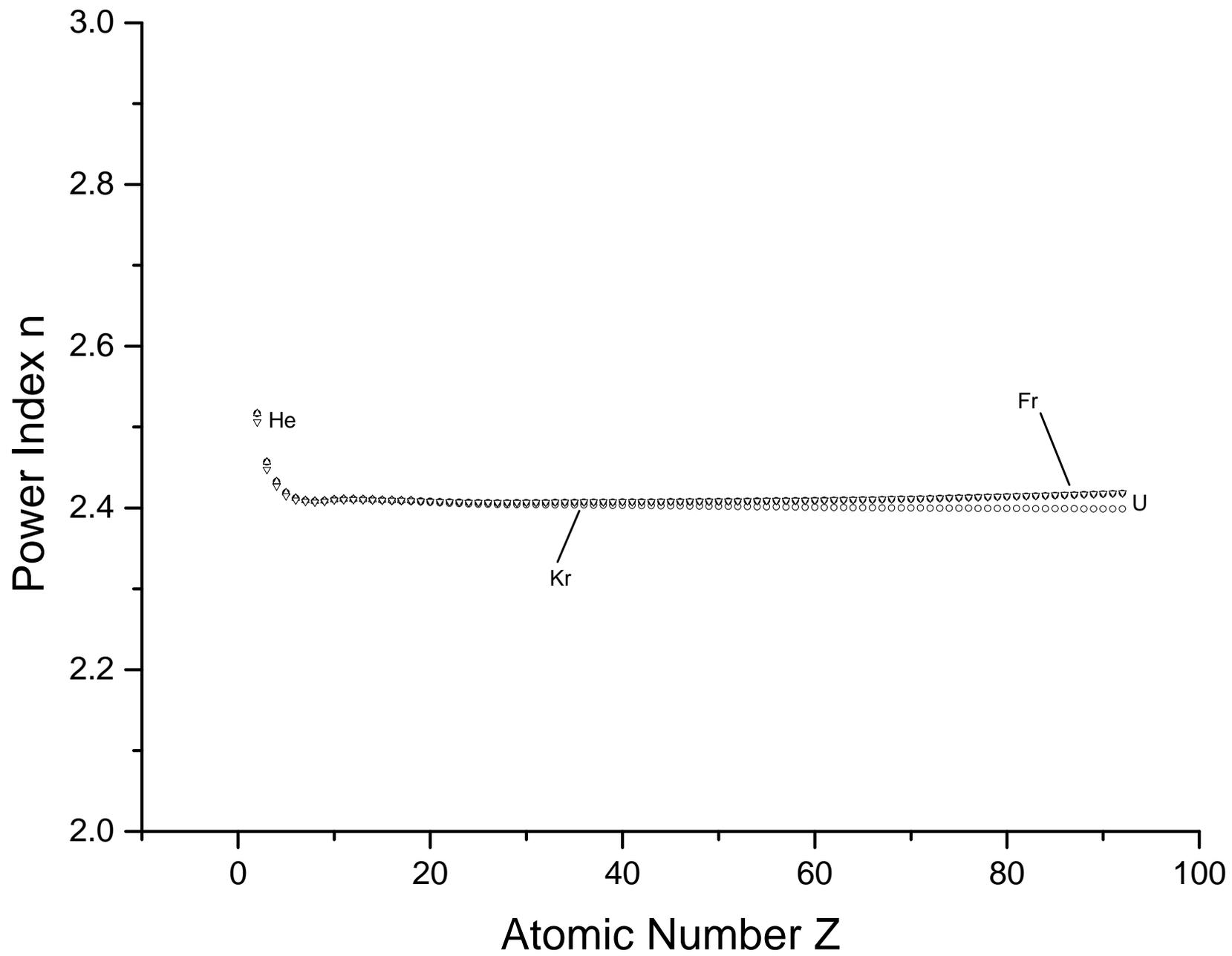
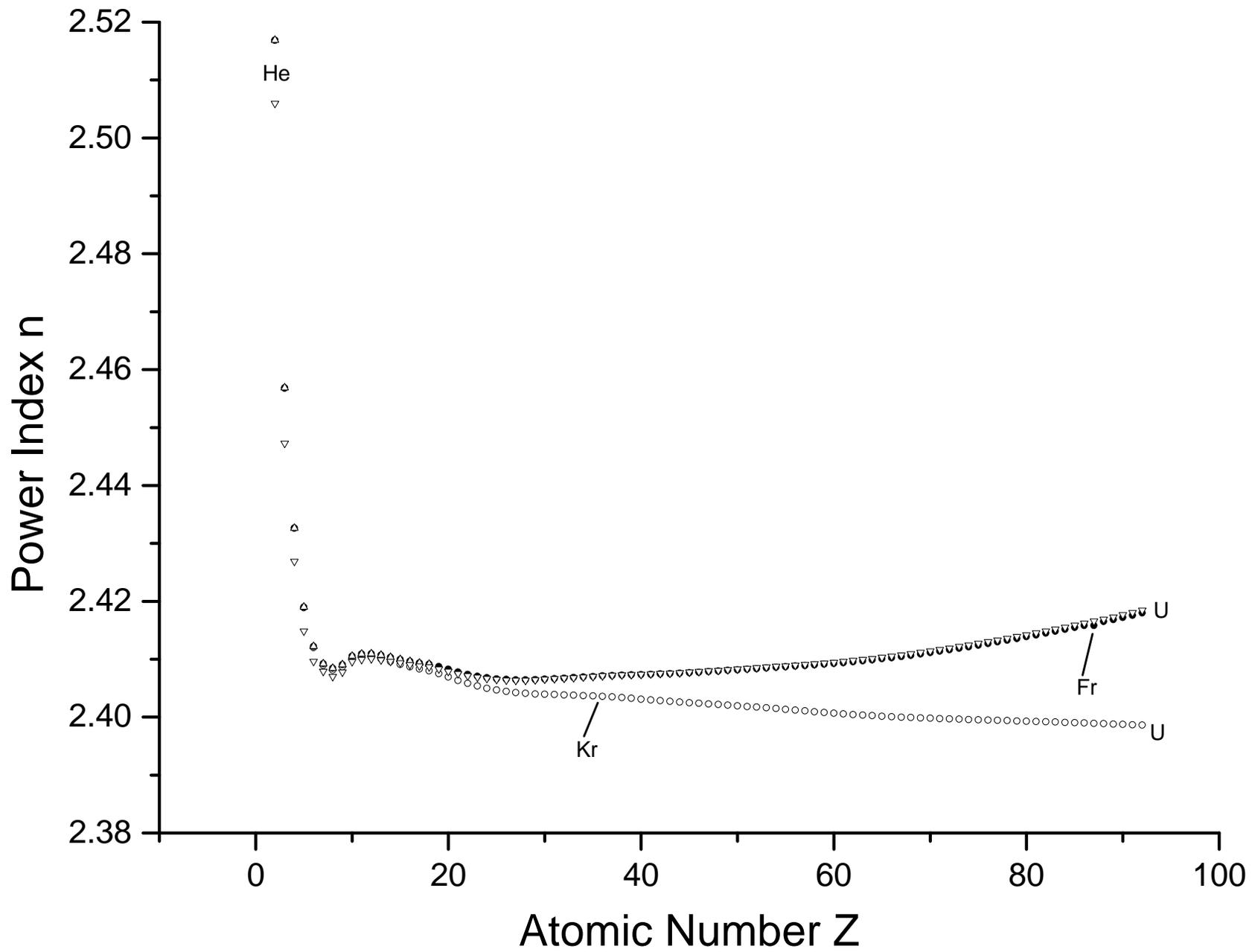
\end{document}